# A Plethora of the Earth-like Planet: Ramifications of a Fuzzy World

Ian von Hegner
Future Foundation Assoc.
Egedal 21
DK-2690 Karlslunde

**Abstract** One primary reason for the formulation of the term Earth-like planet and the search for such planets in the galaxy is because life has arisen in such a world. Thus, this search seems justifiable as it is known here what one is looking for. However, the Earth-like concept represents an attempt to set up sharp boundaries for an inhabited planet, even though nature often comes as continua. The analyses in this work show that the term does not represent a clear-cut entity as a general Earth-likeness cannot be abstracted. Thus, the complex variation of environment and life means that the singular term Earth-like planet is more appropriately treated as a fuzzy world. Such a fuzzification has the consequence of the term being not only more limited than assumed but may even be deceptive, as an Earth-like planet on one hand can be in a segment in which it does not seem particularly Earth-like, but still possesses life, but on the other hand can appear very Earth-like but not possess life anyway. An atmosphere can provide a biosignature by being displaced from thermodynamic equilibrium, derived from antagonistic adaptation, in which life as a double-edged sword, on one hand, continuously makes the external environment less favourable for itself, while on the other, increasing its capacity to do so. Yet, there is an issue with using this as a search criterion for potentially inhabited worlds, as such planets can give impressions that do not reflect what has gone on; they can even give a ghost biosignature. These novel analyses do not represent a limitation in the search for Earth-like planets, as the plethora of Earth-like planets shows the possibility that the number of inhabited worlds can be large, but do represent a limitation in the search for life on such worlds.

## 1. Introduction

When searching for life elsewhere in the galaxy, Earth-like planets are frequently searched for, because life is known to have arisen on such a planet. Thus, one of the primary reasons for the formulation of the term Earth-like planets is an interest in the potential possibility of such planets harbouring life. This seemed justifiable, as searching for life elsewhere in the galaxy involves many unknown unknowns. By rectricting this search to other Earth-like planets, one however restricts oneself to known unknowns. Thus, one circumvents an unknown by focusing on a known, as in this case it is known with certainty that life has arisen on an Earth-like planet.

Life may exist on types of solar system bodies other than Earth-like planets, but here the search has been limited to searching for worlds located in the habitable zone of other stars. The habitable zone is traditionally stated as being the circumstellar region in which a terrestrial-mass planet with a $CO_2$–$H_2O$–$N_2$ atmosphere can sustain liquid water on its surface [Kasting et al., 1993; Kopparapu et al., 2013]. Life on such a world is searched by atmospheric biosignatures, which are a signal life makes to the universe about its presence on a planet.

Although comparative studies of Earth-like planets with regard to life have not yet become a reality, analyses are however still possible, based on the knowledge of one Earth-like planet and the life it harbours.

Astrobiology is the multidisciplinary field that investigates the deterministic conditions and contingent events with which life arises, distributes and evolves in the universe [von Hegner, 2021]. Thus, the subject of astrobiology is not in principle to locate life beyond the Earth, but instead the study of the conditions and events with which life functions and operates on a large scale.

Some issues need addressed in connection with the term Earth-like planet. In the search for life elsewhere, the search is here for Earth-like planets, because life is known to have arisen on such a world. However, that Earth-like planet on which life originated is very different from the Earth-like planet on which life exists today. Thus, the size of the planet and its distance from the sun are the same, but the planet itself was different then from what it is now. So there are basically two Earth-like planets that are quite different over time. That life exists on a planet such as present-day Earth and that life arose on a planet such as the Earth of the past may entail two very different situations.

That a planet changes over time is common knowledge. However, the presence of life on a planet means that this planet can be treated with theory and methods from not only the physical and chemical sciences but also the geological and biological sciences as well, including evolutionary biology, otherwise one has the unknown known, wherein one does not include that which one knows. Thus, the relationship between the changes of a planet and the changes of life can lead to things that are not obvious, as the historical aspect of an inhabited world can mean different rules of the game than for a planet without life, even if this planet is geologically active.

Life as we know it is a planetary phenomenon. It is an environmental phenomenon occurring when natural selection by simultaneous trials acts on chemistry under given favourable conditions over a longer period.

However, life as we know it is also an evolutionary phenomenon. Thus, evolution has a component of local necessity and a component of chance, which will necessarily inform an inhabited Earth-like planet as well.

That the Earth-like planet life arose on is different from the Earth-like planet on which life exists today shows that the term Earth-like planet is unclear to use, as Earth-like in reality implies a plethora of planets over time.

There is also an issue with using the term as a search criterion for potentially inhabited worlds, as searching for an Earth-like planet as the Earth is now with regard to life can entail a different situation than searching for an Earth-like planet with regard to life as it was in the past, or as it will be in the future with regard to life.

These issues will be the focus point in the present work, in which an internalist approach will be taken, i.e. the focus will not be on the habitable zone, but only the world itself and its life will be discussed.

**2. Discussion**
In this paper, Section 3 presents and clarifies the term Earth-like planet and its issues. Section 3.1 discusses that a general Earth–likeness cannot be abstracted. Section 3.2 presents the fuzzy world concept. Section 4 discusses the search for life through its unique characteristics. Section 4.1 presents and discusses antagonistic adaptation. Section 4.2 discusses the deceptiveness of the Earth-like planet in the search for life on it. Section 5 presents and discusses the probabilities that two Earth-like worlds can be similar. Finally, Section 6 summarises the results in this paper and its implications in the search for life in other solar systems.

**3. The Earth-like planet**
When searching for life elsewhere in the galaxy, Earth-like planets are frequently searched for, because life is known to have arisen on such a planet. Thus, one of the primary reasons for proposing the term Earth-like planet is an interest in the potential possibility of life on such planets.

Life may arise on other solar system bodies with other conditions, but it is known with certainty that life has arisen on this planet, and searching for other Earth-like planets therefore seems justifiable. Here we know what we are looking for.

When one says Earth-like, one obviously does not mean exactly a new Earth, e.g. there is the term super-Earth, planets that resemble the Earth, but only larger [Rivera et al., 2005]. Furthermore, the habitable zone does not have to be exactly the same, as the luminosity of stars is known to increase over time, meaning that the habitable zone evolves outwards [Hart, 1978].

However, there is still an issue here, in that the Earth-like planet life arose on in many ways markedly differed from the Earth-like planet on which life exists today.

Thus, the size and distance to the star it orbits is the same, but the planet itself was different then from what it is now.

So given that life exists on both Earth-like planets, but these planets significantly differed from each other over time, there are basically two different Earth-like planets, simplying an issue with the term Earth-like planet.

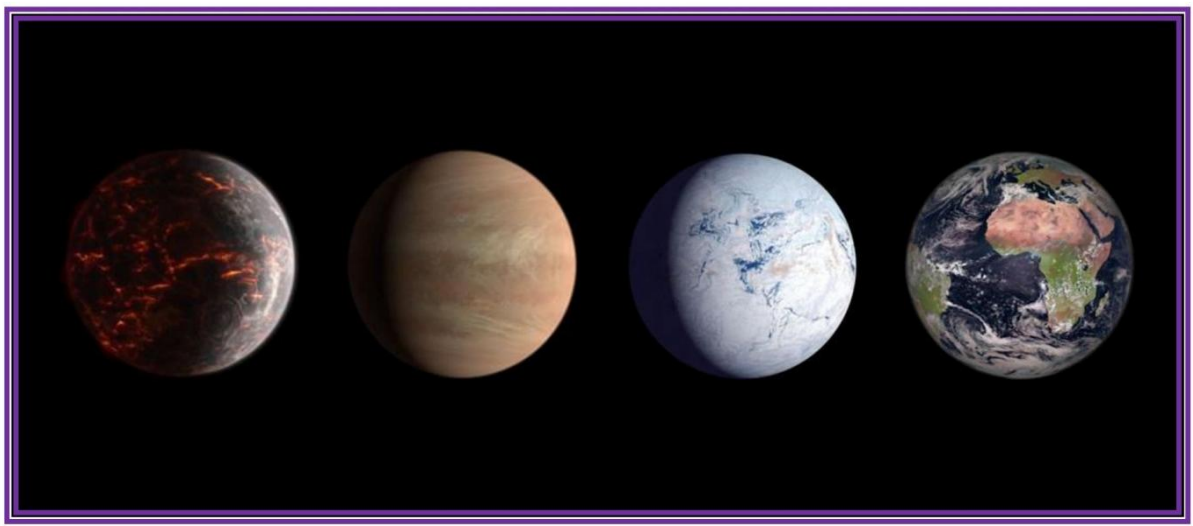

| Hadean | | Archean | | | | Proterozoic | | | Phanerozoic | | |
|---|---|---|---|---|---|---|---|---|---|---|---|
| S1 | S2 | S3 | S4 | S5 | S6 | S7 | S8 | S9 | S10 | S11 | S12 |

**Figure 1.** The plethora of the Earth-like planet. The picture shows several different planets. These planets all have the same size and position in relation to the sun; they are all, in fact, the Earth. The table shows the Earth's eons over time, the Hadean, Archean, Proterozoic and Phanerozoic. It also shows the specific sequence of segments applicable to the Earth, here informed by the Earth's defined eras. A finer division could have produced more Earth types and more segments. An Earth-like planet in another solar system may display other lengths of its eons, meaning that the number of segments for each type may differ from here. Image credits from left to right: Alec Brenner; Keith Cowing; Chris Butler/Science Source; ESA - Earth from Space: Earth Day.

Thus, as seen in Figure 1, it is not immediately obvious what is meant by saying one is looking for an Earth-like planet. All these planets look different, yet they are all, in fact, the Earth. They all have the same size and position relative to the sun. So to single one of them out as being more Earth-like than others seems arbitrary.

As is the case for life, establishing a definition of an Earth-like planet *per se* that everyone agrees on may not be easy.

From a semantic point of view, these different types over time must all be termed Earth-like. Thus, the Earth has always been Earth-like. What else can it be when it is named after itself?

From a purely planetary science point of view, there are traditionally no issues here either. It is known that the Earth has been different over time, but from the semantic point of view, there is not an issue, as all the types are Earth-like, what else could they be?

However, an issue here is that the Earth-like planet on which life arose in the past is different from the Earth-like planet on which life exists today.

Life arose relatively early. Several lines of evidence point to traces of life as early as 4.1 billion years ago [Bell et al., 2015], in the Hadean eon. This eon was, indeed, so different from the Phanerozoic eon that they were, in a practical sense, two different planets.

That life exists on a planet such as present-day Earth, and that life arose on a planet such as the Earth of the past may entail two very different situations. If an Earth-like planet is located in the habitable zone around another star, it may appear very different from present-day Earth. Thus, searching for an Earth-like planet as the Earth is now with regard to life may entail a different situation than searching for an Earth-like planet as it was in the past or as it will be in the future in the same regard.

It all comes down to the very statement of the term Earth-like planet, a term which may not be justifiable.

*3.1. Variance*

The fundamental assumption underlying the term Earth-like planet is that it is possible to assign value to it, i.e. that it is possible to assign value to an abstraction of the Earth.

However, as mentioned, it is not immediately obvious from Figure 1 what is meant by the term Earth-like planet. All these planets look different, yet they are all, in fact, the Earth. They all have the same size and position relative to the sun.

Yet, the use of the term Earth-like planet assumes the existence of a 'general Earth-likeness', which exists as a discrete quality throughout the history of the Earth. It is an abstraction of the Earth as a single entity. However, the Earth is not a thing frozen in time, but a variation over time. Thus, there is an issue with using an abstraction of Earth-likeness as a single entity.

The fundamental assumption underlying the term Earth-like planet is that it is possible to assign value to it, i.e. that it is possible to assign value to it by measuring Earth-likeness as a single quantity.

One collects a large number of data that together will produce Earth-likeness. Thus, a result for the present-day Earth-like planet measures the general Earth-likeness because general Earth-likeness is defined as a measurable thing given specific data.

Thus, data for the present-day Earth-like planet measures the general Earth-likeness of the Earth because general Earth-likeness is defined so that certain questions, and the responses to them, will form a cluster of Earth-likeness. Thus, data for the present-day Earth-like planet will yield a certain result. However, the Earth has been different over time, meaning that the Earth will generate different data at different times, and thus create a different compilation of data and thus a different result. Thus, a past or future Earth will not possess general Earth-likeness. This means that different questions, and thus the different answers to them, will generate different results. This demonstrates that Earth-likeness can be properly considered a combination method, which utilises distinct examinations of distinct things.

So even though a quantity designated the present-day Earth-like planet is a measurable thing, it does not *ipso facto* mean that general Earth-likeness is genuine. Thus, ultimately one can question the validity of the term general Earth-likeness.

A dynamic planet such as the Earth is marked by complex variation over time. This variation is continuous. So focusing on a snapshot of it and conferring representative status on it does not accurately represent what is going on. In evolutionary biology, one does not focus on individual points, on the mode or the mean on a graph, but on the variation over time. Thus, life is not a thing frozen as a snapshot in time but is a variation over time.

Thus, singling out one of these planets in Figure 1 and its life as being more Earth-like than others seems arbitrary. The Earth is not a frozen thing, a single number; thus, converting the entirety of the Earth into single entities does not capture the complex history that the Earth and its life represent. So the Earth is not covered by a ranking, that is, an ordering of its complex variation as a gradual ascending scale, as there are no Earth types that are more true than others with regard to life or geology.

One cannot therefore meaningfully abstract a general Earth-likeness amidst the variation, achieve a quantification as one number for that planet, and rank the Earth-like planet in a single series with different values.

Although a general Earth-like planet cannot be abstracted, the difference between the past Earth and the present Earth is still significant in terms of life, and thus significant for astrobiology in the search for such planets. Thus, the term Earth-like planet with regard to the search for life elsewhere must be used more specifically than it is being done now.

The term Earth-like planet implies a singular form, but the Earth is in reality in plural form, a plethora of planets over time and space. So rather than operating with just a single Earth-like planet, one can benefit from slicing the Earth into several segments over time.

The Earth is 4.56 billion years old [Bouvier and Wadhwa, 2010], and geologically this period has generally been divided into 4 eons and 10 eras [Cohen et al., 2023]. Thus, in the following sections, the Earth for the sake of argument will be divided into 12 segments, as seen in the table in Figure 1. The Hadean eon will here be divided into two eras, as life or chemical evolution probably occurred in segment S2. All but the first and possibly the second of these segments possessed life.

None of these Earth-like segments are truer than others. Thus, present-day Earth in S12 is no more representative or even real over time than others, just as present-day life is also no more real or representative of life over time. So the segment S2 with the first life differs from the current segment S12. However, segment S3 where life spread is different from segment S11 before the current segment S12 etc. They are all different and none more true than others.

One could perhaps say that Earth-like is the sum of all these segments, but that is not accurate either, as this will simply be an abstraction of a new segment that is not representative of the individual segments, and thus not of the Earth over time.

All this is perhaps best illustrated through the world line of the Earth. The Earth is usually presented in a two dimensional Copernican-Keplerian diagram of the solar system. In this diagram, the sun is represented as a large circular disk, and the Earth as a smaller disk near it. The orbit of the Earth is displayed as a dashed ellipse on the flat page. This representation however captures only one instant of time. In the Einsteinian-Minkowskian presentation, the picture is more complicated. Thus, the Earth is not merely a sphere but also, in fact, a helix, a long filament spiralling around the sun, from the latter's past to the future. This filament or helix is the planet's world line through space-time. All true objects possess four dimensions: width, breadth, height, and duration. Thus, the Earth has an extension in time.

Seen in such a world line, the Earth is not a fixed geological constant. It is never a clear-cut entity, but appears as a fuzzy dynamic entity varying in space-time. The various snapshots or segments on the Earth's world line are thus each a frame in which the Earth's life exists in interaction with the environment.

Of course, the 12 segments do not stop in this 4-dimensional space-time diagram with the present-day Earth-like planet. The Earth will continue to change into new different segments after S12 until changes in the phases of the sun towards the red giant stage will end life on the Earth, in fact, probably end the Earth itself.

It even applies that the planets in a solar system drift away from their star. Thus, the orbit of the Earth does not stay the same over time, but instead spirals slowly outward. Even the moon moves away from the Earth over time, making the Earth-like planet as defined in relation to its star even harder to justify.

This division into segments is not a division in complexity, but simply the Earth's extension in time. Thus, the contemporary segment S12 is not more complex than the previous segments; it does not represent the Earth distributed over all its segments. So the complex variation distributed throughout the Earth's history, with regard to life, is still there.

One could say that the segment in which life arose, which may be S2, was special compared to the other segments before and after. Yet, segment S1 before S2 led up to the building of the first life; life does not just arise in an instant but is built up over time. Thus, one cannot take a complex variation and treat a point in it as one number and abstract the emergence of life as a single entity. Nature frequently does not come in the form of clear-cut essences and definite boundaries, such as a clear-cut beginning of life, but rather in the form of continua. So just as life in the later segments has been in continuous variation, so the build-up of life in the first segments has continuously been there.

*3.2. Continua*

The Earth-like concept represents an attempt to set up sharp boundaries for a planet and its life, even though nature often comes as continua.

Thus, it may appear contradictory here on one hand to point out that nature comes in continua, that one cannot rank the Earth on an ascending scale, but on the other, divide the Earth-like planet into segments. However, the history of humankind, in fact, the history of all life, can illustrate the deeper meaning behind this.

For approx. 4–8 million years ago, a group of nonhuman primates lived in Africa [Bradley, 2006]. These had children who did not significantly differ from their parents either in genotype or phenotype. This situation continued. In generation after generation over millions of years, no children were born who differed significantly from their parents.

Yet! 4–8 million years ago, there lived a group of nonhuman primates, and today there exists a group of human primates who call themselves *Homo sapiens*.

It is a descent that has undergone modest change throughout each generation, and this fuzziness between the boundaries of each generation creates the impression that the species had remained the same through time. Yet, if fitting geometries of space-time made it possible for either the nonhuman primate or the human primate to cross time and meet each other, then it would almost certainly not be possible for them to cross-reproduce, because they are in fact not the same species.

There has not been a deterministic path between nonhuman primates and human primates. While the human primate in retrodiction can trace its origins back, the nonhuman primate could not predict the emergence of the human primate. Thus, there have been many different outgoing branches from the group of the nonhuman primate, many species, which means that all of these except one have disappeared again, displaying a fuzziness here as well.

There is a crucial insight here. There are fuzzy boundaries between each generation here. Yet, species at each end of this continua markedly differ from each other. The same applies geologically, where environmental changes occur slowly through fuzzy boundaries, yet at each end of the segment, there is a marked difference. There were no sharp boundaries between the segments, each subsequent segment followed almost as a continua from the previous one. And yet, in the beginning, there was the Hadean eon, and now there is the Phanerozoic eon, which markedly differ.

So these fuzzy boundaries mean that it is not possible to point to a specific place and say that here there was a definite boundary between the nonhuman primate and the human primate, or that here humanness arose. Similarly, it means that it is not possible to point to a specific place and say here a segment is more true than the other segments, or that here one can abstract a general Earth-likeness quality among them.

Thus, the Earth-like concept represents an attempt to seek sharp boundaries. However, the complex variety of life and the complex variety of environmental changes makes the singular term Earth-like planet more appropriate when treated as a fuzzy world. Such a world does not deal with a simple binary yes or no choice, but with fuzzy measures and boundaries.

Such a fuzzification of the Earth-like planet, however, does not easily elicit clear answers to the presence of life when searching for it.

**4. The search for life on Earth-like planets**
The way to find Earth-like planets is per definition to find them in the habitable zone around a given star. As discussed, one cannot posit the existence of general Earth-likeness. However, if one cannot posit such a thing, then it is not easy to set up a search criterion either.

But if one wants to find inhabited Earth-like planets, the method is to exploit certain unique characteristics that life exhibits. Thus, in addition to the fact that environments can change over time, it is also a fact that life can change the environments it exists in. This can occur even on a planetary scale, as demonstrated by an atmosphere far from thermodynamic equilibrium.

In an ecological setting, organisms can be said to exist in equilibrium with other organisms, and with their environment. Such an equilibrium is always short-lived, however, a truce that is eventually broken; yet for a time it exists. However, in a biophysical setting, there is no equilibrium between life and the environment but only a persistent disequilibrium in the setting.

Thus, a cell always increases the entropy of the surrounding external environment, in the form of heat and waste products, while it lowers its own internal entropy, in the form of maintaining order and function [von Hegner, 2021].

Thus, from the insight that life increases entropy in the environment, an atmosphere far from thermodynamic equilibrium on a planet was proposed to indicate the presence of life on that planet, as this atmosphere far from thermodynamic equilibrium will be caused by life [Hitchcock and Lovelock, 1967]. Thus, an atmosphere can provide a biosignature by being displaced from thermodynamic equilibrium, implying the presence of large amounts of life.

However, there is an issue here in that it is possible for abiotic factors under certain circumstances to show such an effect as well, which is why this approach has since been done with caution.

Yet, while it is agreed that abiotic factors may well produce an atmosphere far from thermodynamic equilibrium, it is still assumed that a biotically produced atmosphere far from thermodynamic equilibrium must be present if life is present on an Earth-like planet.

Thus, the present-day Earth is affected by its life, its atmosphere is far from thermodynamic equilibrium, so technically one would think that a planet without life cannot be Earth-like. However, some deeper issues here are relevant as well to the Earth-like concept. Because while this method is indeed valid if purely chemical factors can be eliminated, and it can thus give the results it promises, it however examines only one segment over time out of many with Earth-like planets.

It is a practical example that focusing only on a segment designated as general Earth-likeness is not appropriate if, when searching for inhabited planets, one focuses only on planets with an atmosphere far from thermodynamic equilibrium, because that an inhabited Earth-like planet must present an atmosphere far from thermodynamic equilibrium is not correct.

Thus, contrary to the original idea behind an atmosphere far from thermodynamic equilibrium, the absence of this is not an inevitable sign of the absence of life, at least in all segments of an Earth-like planet. After all, an atmosphere far from thermodynamic equilibrium is a by-product of life but not initially a prerequisite for life. Thus, the first life evidently existed without such an atmosphere [von Hegner, 2020]. Thus, the fact that a planet does not exhibit such an atmosphere far from thermodynamic equilibrium may have other reasons than simply the absence of life.

An inhabited Earth-like planet that does not have an atmosphere far from thermodynamic equilibrium may be a young planet where life has not existed long enough to have brought the atmosphere out of equilibrium. In the segment where the first life arose and the segments where life began to spread, life had evidently not existed for that long. Thus, that the atmosphere far from thermodynamic equilibrium is not there indicates here that life existed there only for a short time and thus did not have the time to affect the atmosphere.

An inhabited Earth-like planet that does not have an atmosphere far from thermodynamic equilibrium may be a young planet in which life does not exist in sufficient numbers to have brought the atmosphere out of equilibrium. In the segment in which the first life arose and in those in which life began to spread, life had evidently not existed in large quantities. Thus, the absence of an atmosphere far from thermodynamic equilibrium indicates that life exists only in limited numbers and thus did not have the amount to affect the atmosphere.

An inhabited Earth-like planet that does not have an atmosphere far from thermodynamic equilibrium may be a planet in which life exists in large quantities, but has not yet been able to affect the atmosphere. Thus, even a sufficiently large number of lifeforms will require time before they can change something as large as the atmosphere of a planet.

Thus, life may have existed for a long time on a planet in large quantities but has not yet have been able to affect the atmosphere. Thus, this search method suits the later segments of the Earth's history, in which the Earth exhibits an atmosphere far from thermodynamic equilibrium, but does not suit the Earth's earlier segments, in which the Earth did not exhibit an atmosphere far from thermodynamic equilibrium, but in which life nonetheless existed. This is important to emphasise, as the Earth has had life for almost its entire history.

An inhabited Earth-like planet that does not have an atmosphere far from thermodynamic equilibrium may be an old planet where life is not successful. Thus, on this planet, life has existed there almost from the beginning, but has not been successful and does not exist in large numbers. Species come and go, and even a species that has managed to survive over a considerable period, but is a sole representative of a once widespread order, is not considered successful from an evolutionary biology point of view [Gould, 1991]. Thus, that life succeeded in arising in the previous segment differs from a situation in which life will subsequently succeed in the present segment.

An inhabited Earth-like planet that does not have an atmosphere far from thermodynamic equilibrium may be an old planet where life only arose late. Thus, one could expect a direct proportionality between a planet's age and its biosignatures, i.e. a young planet has a low amount of life and thus no atmosphere far from thermodynamic equilibrium, and an older planet has a lot of life for a long time and therefore an atmosphere far from thermodynamic equilibrium. This direct proportionality seems to apply to the Earth. However, there seems to be no compelling reason to assume that life cannot arise later on a planet. Thus, an old Earth-like planet can have life that is young, meaning that it is not only in the first segments that there can be a low amount of life, and thus no atmosphere far from thermodynamic equilibrium, but also in the later segments on an Earth-like planet. The expected proportionality can therefore be deceptive as a search criterion.

Thus, here it can be seen that this search method even in the later segments of an Earth-like planet's history is not necessarily suitable, as the Earth-like planet does not present an atmosphere far from thermodynamic equilibrium, but where life nonetheless exists. This is important to emphasise, as the age of a solar system does not necessarily indicate which of the segments one is in here and now. So an Earth-like planet in each solar system can be in different segments with regard to life, even though these planets are the same age.

But despite all this, the present-day Earth-like planet exhibits an atmosphere far from thermodynamic equilibrium produced by the presence of life over a long period in large quantities, so this is, under these given circumstances, a valid approach. However, while valid, the issue is that here one focuses on a single segment that shows an effect that each of many of these segments do not.

The issue is more precisely that the segments that show an atmosphere far from thermodynamic equilibrium cannot be said to be more important than the others, as this segment no more represents the entire history of an Earth-like planet than the other segments do.

One can say that a quantification is carried out here with one number for a planet in which one ranks planets according to this atmospheric effect. However, one cannot rank these segments by importance, as they all possessed life. In the light of the examples reviewed, it can even be debated whether the present-day Earth represents the minority of inhabited contemporary Earth-like planets.

If the purpose is to search for life on an Earth-like planet with an atmosphere far from thermodynamic equilibrium, then this is of course the right approach, but it will necessarily equal a fuzzified approach.

As reviewed, an Earth-like planet is conducive for life detection only in certain of its segments. Thus, focusing on a single segment that shows an effect that many of these segments each do not show and on that basis abstracting a single entity among all the segments called general Earth-likeness, is not valid. Earth-like is all inhabited segments over time, and is not given by singling out one of them.

Thus, the Earth is a fuzzy world, meaning a world where the presence of life cannot always be determined unambiguously. It does not give an easy clear-cut answer to the presence of life on that planet in a search.

*4.1. Antagonistic adaptation*
A segment with an atmosphere far from thermodynamic equilibrium is, as reviewed, not representative of the entire history of an inhabited Earth-like planet. However, entropy displacement indeed represents life, and that is important. Thus, an atmosphere far from thermodynamic equilibrium represents a global survey of a potentially habitable planet, while a potentially habitable planet without such an atmosphere far from thermodynamic equilibrium would represent a local survey.

The fact that life does not manifest itself in an atmosphere far from thermodynamic equilibrium means that searching for life in this way is not possible by observing the atmosphere from a distance. However, it will still be possible if one can do the study on that planet itself, as life cannot avoid leaving traces locally due to entropy displacement, and will thus always leave a signature. Thus, life exhibits certain unique characteristics, and even in this situation, entropy displacement plays a role, as antagonistic adaptation will be in effect.

As mentioned in section 4, a cell always increases the entropy in the surrounding external environment, in the form of heat and waste products, while it lowers its own internal entropy, in the form of maintaining order and function. Thus, a cell unavoidably makes its surrounding environment less favourable for itself [von Hegner, 2021].

In that sense, an organism does not exist in harmony with the environment but exist displaced from thermodynamic equilibrium. It will continuously and unavoidable make the environment less favourable for itself in order to live.

In a thermodynamic interpretation, evolution can be said to be a competition to best utilise the available energy [Lotka, 1922a,b; Odum and Pinkerton 1955].

It can also be said in an interpretation that life forms become more complex through evolution. By becoming more complex, they also become more effective at increasing entropy in the external environment even faster.

A cell is an entropy increaser. Thus, the more complex and efficient a life form becomes, the more entropy it will produce. Thus, a direct proportionality is at play here; the more effectively adapted a life form is to capture and transform free energy, the more it will increase entropy, and the more it will make its surrounding environment less favourable for itself.

Thus, there is a double-edged sword here, as a life form on one hand continuously makes the external environment less favourable for itself, while on the other, increasing its capacity to make it less favourable for itself.

So the apparent paradox is that while a life form through adaptation becomes fitter in terms of utilising available energy, it also implicitly brings about its own doom as it also becomes better at generating more entropy in the environment.

Thus, there is a direct proportionality between the increased adaptation to energy conversation and increased entropy production. This means that there is an inverse proportionality between increasing the ability to live in that particular environment and simultaneously decreasing the ability of that environment to support that life.

Thus, in one interpretation, it could be stated that an inherent flaw exists located in the very operation of life itself, as the two driving principles of life, energy conversion and adaptation, basically work against each other.

Thus, a deep conflict exists at the most fundamental level between the two very ways life must necessarily function.

This is not only for life as we know it, as all work entails energy conversation, and entropy increase is thus inevitable in this universe. Thermodynamics does not impose this restriction on life; rather, it is an innate feature of the way the universe works, and a cell will therefore unavoidably make its surrounding environment less favourable to itself.

Life overcomes this apparent paradox by locally moving to new environments, and by globally existing in a vast open system, i.e. the Earth, and ultimately, the universe. Thus, for all practical purposes, this restriction can be ignored, but the more restricted their environment is, the more it will be noticeable. On a planet such as the Earth, the effect of individual cells in a larger environment is negligible, and life can handle it at the organismal level, but the total atmosphere can however eventually experience the effect of this.

An example of when life cannot always bypass the effect of making its surrounding environment less favourable for itself is the Great Oxidation Event (GOE) in which the Earth's atmosphere and its shallow ocean underwent a rise in $O_2$ between 2.4 and 2.1 billion years ago [Holland, 2002; Lyons et al., 2014]. The GOE may have been due to cyanobacterial activity, whereby oxygen was produced as a waste product, which accumulated in the Earth's atmosphere, eventually significantly changing the Earth system [Cloud, 1968; Lyons et al., 2014]. This rise in $O_2$ may have triggered severe extinction events for the many then-existing anaerobic species [West, 2022], and although this provided opportunities for new adaptations, the point is that although the pollution of individual organisms was negligible, the organisms overall made their surrounding anaerobic environment less favourable for themselves, and eventually brought about their own destruction.

Yet, compared to these interpretations, it can be said that the increase in entropy is a constant, since with any kind of work, energy will be converted.

But the increase in complexity is a variant, in that life can move towards increased complexity or it can move the opposite way.

Evolution is fundamentally variation. It starts as a lump of variation spreading outwards in all directions. There is no direction and no goal for the spread of variation, and most of the variants disappear again in competition with each other and against the pressure of the environment. However, a few survive, and they themselves become a lump from which variation spreads outwards in all directions. There is no drive towards the variants becoming more complex. They simply adapt to changing local environments.

Thus, life can move towards complexity, but can also move towards simplicity, which is seen in e.g. reductive evolution, in which the number of genes is reduced, or anatomical, in which life forms eventually become parasites. Thus, if the environment is accumulating waste products, then life as an adaptive response can in effect move towards becoming simpler, i.e. less efficient in utilising available energy. Thus, in a population of organisms, entropy production can switch between being turned up or down over time, depending on the complexity of their adaptations.

That life is governed according to a maximum power principle [Lotka, 1922a,b; Odum and Pinkerton 1955] is also a debatable interpretation. Thus, evolution is adaptation to changing local environments. It can work only with what it has. Natural selection selects between the variants in a population that ultimately arose by mutation. These may be the best choice in the given local environment, but are not necessarily the optimal design for energy efficiency, from a bioengineering point of view.

But despite all this, it still applies that life leaves its mark locally by increasing the entropy in the surrounding external environment, in the form of heat and waste products, while it lowers its own internal entropy, in the form of maintaining order and function.

Thus, in that sense, life is not a beneficial natural phenomenon, as other natural phenomena do not through teleonomic activities (those directed by a program of coded information, e.g. the organism's own hereditary material) go against entropy by lowering their internal entropy at the expense of the surroundings. Other natural phenomena proceed instead through teleomatic processes (those resulting from physical laws).

That life makes the surrounding environment less favourable for itself does not pose an issue on a planet such as the present-day Earth, partly because life circumvents this very successfully and partly because it creates an atmosphere far from thermodynamic equilibrium, usable with regard to the search for life.

Thus, the advantage is that life even on a world that lacks an atmosphere far from thermodynamic equilibrium cannot hide; it always leaves traces, even on completely different segments in the past Earth-like planet and the present Earth-like planet, but the disadvantage in this situation is that it will be necessary to investigate locally, i.e. on the planet itself.

*4.2. The fuzzy world*
The detection of life through an atmosphere far from thermodynamic equilibrium leads back to an Earth-like planet, but an Earth-like planet does not necessarily lead to the possibility of a detection of life through an atmosphere far from thermodynamic equilibrium. Thus, the reviewed examples in the previous sections showed that a fuzzy world can be deceptive in all these ways in terms of searching for signs of life on it. It can in fact be even more deceptive than this.

If all life disappears relatively quickly from the present-day Earth-like planet, perhaps due to some hypothetical external global event, then the influence of life on the atmosphere will disappear. However, the atmosphere will not reach thermodynamic equilibrium immediately, as some time will necessarily pass before this happens. Thus, interestingly, a 'ghost biosignature' will appear for life that is no longer there on a planet, demonstrating the fuzzy boundaries between life and biosignatures, and a deceptiveness regarding the presence of life. So here it will be some time before the atmosphere of the planet returns to the state it had before life was able to affect the atmosphere, whereby a new segment in fact exists for a time on this planet.

If life slowly disappears from the present-day Earth-like planet, then life's influence on the atmosphere will eventually disappear. However, the Earth will not reverse back to the segment it had when there was no life, because while the Earth has had life for almost its entire history, life arose at the end of the Hadean eon, and the Earth will not return to this particular eon again. So here a new segment will eventually arise that has not previously existed on the Earth. This segment may show a geologically dynamic, yet relaxed planet where life could be expected; in fact, if life was transported there, it could possibly live there. However, as there is no life there, one can be led to believe that it has never been there, as one would expect an Earth-like planet of the same age as the Earth, which is so fertile that if it had the opportunity, then it would have arisen. Thus, in that way, a fuzzy world can be deceptive in that it does not accurately reflect what has gone on.

If life disappears from the present-day Earth-like planet, quickly or slowly, and a new segment arises, then life may not arise again, as this new segment differs from the segment in which life arose, and even if this new segment is such that life from an Earth-like planet such as present-day Earth or the previously inhabited segments could probably live there if placed there.

There is also the fundamental issue of whether two Earth-like planets can both be genuine Earth-like, if one possesses life and the other does not, even if all the solar system and geochemical conditions correspond to each other.

Strictly speaking, if one specifically looks for a planet such as present-day Earth, which has no life, then one would opinionate that it cannot be an Earth-like planet, as the present-day Earth is affected by its life, whose presence has been a constant through almost the entire history of the Earth.

But as discussed in the previous sections, an Earth-like planet without life may well display an atmosphere far from thermodynamic equilibrium due to chemical factors reminiscent of the biosignature displayed by present-day Earth. Thus, this issue depends on whether one solely defines an Earth-like planet as having a certain size and existing in the habitable zone around a star, or whether the presence of life on it is a requirement. Thus, this is a question open to debate, but from the search point of view, they are both Earth-like planets, as in the mentioned cases one cannot remotely detect a difference between them.

Thus, an Earth-like planet represents a fuzzy world, not because the search for life in the mentioned ways is not a sound approach, but because this variation of life and environment brings an inherent fuzziness to such a planet.

This fuzzification has the consequence that the use of the term Earth-like planet may be more limited than assumed. It may even be deceptive, as an Earth-like planet on one hand may be in a segment in which, to an external observer, it does not seem particularly Earth-like and yet possesses life, but on the other, it can seem very Earth-like, yet not possess life.

The changing segments, the complex workings of life and interaction with a complex environment mean that such planets can produce impressions that do not reflect what has gone on. They can even create a biosignature that is not due to the existence of life or to abiotic factors, but to a time delay.

Thus, an Earth-like planet can in certain segments have the same appearance regardless of whether there is life on it.

So Earth-like planet is limited here as search criteria. The fuzzy measure is not necessarily due to insufficient search technology, but to the very nature of life on a world, and in fact, to the dynamic world itself. A segment may seem different from today's segment, but still have life. Another segment may appear such as the present-day segment, but have no life.

So searching for an Earth-like planet as the Earth is now entails a different situation than searching for an Earth-like planet as it was at some time in the past, as it will be in the future, or as it will be in another solar system. This point is important to address, as a diversity of Earth-like planets also implies different approaches.

As mentioned in section 4, even the age of a solar system is not necessarily indicative of which of the segments a planet is in here and now with regard to life. So an Earth-like planet in each solar system can be in different segments with regard to life even if these planets have the same age.

It is debatable whether the Earth could have moved forward only the way it did. Would the Earth's physical, chemical and geological parameters have progressed the way they did no matter what? Thus, if another Hadean Earth-like planet proceeds to be another Archean Earth-like planet, will it necessarily proceed to a Snow ball Earth-like planet, or can it proceed to a present-day Earth-like planet without this intermediate step? Thus, will a fuzzy world extend to be a planet in which one cannot predict the order of all the segments in relation to each other, even if two Earth-like planets have the same age?

It is clear, however, that a planet with physical, chemical, geological and biological parameters does not move forward deterministically, as life and environment influence each other, and given that evolution has a component of local necessity and a component of chance, it does not follow a predictive deterministic path.

Thus, segments with regard to life will not necessarily be the same between two planets, and even segments without life may be different between two planets. Thus, searching for a planet based on the age of that planet will not suffice.

This fuzziness may also be in effect in other solar system bodies that are not directly Earth-like through their size or location in a habitable zone.

Thus, these discussed scenarios may also hold implications for planets such as Venus. This was once a more environmentally relaxed world, which due to a runaway greenhouse effect, became the inhospitable planet it is today. Life may have arisen on it. However, as the planet's conditions for life worsened, life may gradually have withdrawn into the cloud layer [Haber, 1951; Greaves et al., 2021]. Thus, life may still exist in Venus' cloud layer, but in such modest varying amounts that it is unable to affect the atmosphere so much that a clear-cut biosignature can be detected. Thus, biosignatures can, so to speak, blink in and out of existence over time on that world.

These discussed scenarios may also have implications for planets that are not Earth-like in size. Thus, some of these scenarios are reasons why there could potentially be life on present-day Mars, even though an atmosphere far from thermodynamic equilibrium is not seen there. Mars was in its Noachian system a more hospitable place for life than it is today [Fastook & Head, 2015; Carter et al., 2010], and life may have originated on it. Life on that planet may thus be ancient, but the amount of this life may have gradually shrunk as the conditions on the planet with regard to life worsened. Thus, the influence of life on the atmosphere would eventually have disappeared, but life may still be there.

Thus, worlds such as both Venus and Mars may at certain times have had segments much in common with those in the Earth's history, which is therefore not dependent on being in the habitable zone.

On such planets, it may also become an issue that life continues to make the surrounding environment less favourable for itself. On the Earth, the fact that life makes the surrounding environment less favourable for itself will not pose an issue, as life circumvents this very successfully by moving to new environments, displacing it to the atmosphere, or life is so diverse that some of it can break down the waste products.

On present-day Mars, a few subenvironments suitable for life may still exist. However, these remaining sites are both semi-shielded from the rest of the planet's harsh and probably also ever-shrinking environments. Thus, here it may pose an issue, as the very environment in which life can live will be polluted by waste product accumulation, unable to escape, and the limited habitable environment means that only a limited diversity of life can exist, and may thus not produce life forms that can break down these products. So the very way life works may be the reason that life ultimately does not exist on a planet such as Mars.

**5. The probabilities of Earth-like planets**

One reason to search for another Earth-like planet is, as discussed, its potential possibility of possessing life similar to that of Earth. However, an issue here is that two Earth-like planets, even if they share the same initial conditions, subsequently entail different situations forward in time. To analyse this, one can, for the sake of argument, apply a pseudo-retrocausality analysis to these situations.

Thus, to randomly change the present-day Earth-like planet, only small changes need to be applied in one of the distant past segments to adjust large events in the present segment.

However, if one wants to observe one small specific event in the present segment, one has to apply massive changes in one of the distant past segments to obtain it.

Thus, this means that to change another inhabited Earth-like planet so that it is completely like present-day Earth, one has to apply massive changes in one of the distant past segments to obtain one small specific event in its present segment. Life can go in many directions, and the current direction on the Earth is just one contingent option among many. Thus, the presence of a specific biodiversity represents from an evolutionary point of view only one very small event.

This analysis approach is relevant as an illustration because the fact is that if two Earth-like planets share the same conditions in the present, then this will have entailed different situations in the past.

Thus, the term Earth-like planet by its very definition comes with a built-in limitation because searching for a planet that is exactly the same as present-day Earth in terms of segments and biodiversity is to search for a planet encompassed by several extreme improbabilities.

Thus, if two Earth-like planets share the same conditions in the present in terms of segments and biodiversity, then this will have entailed several highly improbable situations for one of these planets in the past. This situation means that the backward path of life on completely similar inhabited present-day Earth-like planets represents an increasing improbability back in time.

A kind of inverse proportionality applies here, in that the more similar two present-day Earth-like planets are with regard to segments and biodiversity, the more unlikely they are.

This means that life on two Earth-like planets from the segment in which life originated will follow different courses through the segments to arrive at present-day Earth-like planets. So in order for these two present-day Earth-like planets to be completely alike with regard to segments and biodiversity, then life must have followed the same course through all the segments; in fact, the segments themselves must have followed the same course on the two planets.

However, the fact that life by itself follows different courses means that the probability of biodiversity being exactly the same on two Earth-like planets is essentially non-existent. Thus, the very dynamics of an inhabited Earth-like planet prevents two Earth-like planets from being exactly alike in terms of biodiversity.

Of course, the two Earth-like planets are encompassed by an unconditional probability in relation to each other, but it still applies that the segments and biodiversity of another Earth-like planet proceeding in the same sequence as the Earth and becoming completely identical to it entails several extreme improbabilities.

One may think that if one encounters an Earth-like planet with the same age as present-day Earth, then its biodiversity, due to its long history, would have achieved the same form as the Earth's. However, evolution is adaptation to changing local environments. This process has no long-term direction or goal. Today's terrestrial species do not exist due to determinism but because all other species they shared the planet with have gradually disappeared, and they themselves have remained here until some other species replaces them.

Thus, if this variation of life from the beginning of life to the present on two Earth-like planets should be exactly the same, having followed the same course through all the segments, then this would entail an extreme improbability of having arrived at a segment such as the present, rich in the same amount of biodiversity.

So forward or, as discussed here, backwards, the order of the segments and the biodiversity they harbour between two Earth-like planets will not be the same due to an inherent fuzziness for the direction of life.

One may also point to convergent evolution as a counter to this. However, convergent evolution, that is, the creation of analogous structures in different species, does not come to the rescue here, as this does not represent determinism. It does not create the same species, but merely an approximate adaptation of the same design to certain types of environments. There are still many paths for life through the segments, and the segments themselves need not be the same. Thus, the statistical improbabilities exceed, by far, the approximate adaptation.

Thus, summa summarum, that the order of the segments is exactly the same and that the biodiversity is exactly the same on two inhabited present-day Earth-like planets represent an increasing improbability back in time for one of these planets.

So taken to its extreme consequence, this situation means that the increasing improbability back in time will eventually move past the point where life could even arise on the one Earth-like planet, meaning that life cancels out, so to speak. Thus, following this line of thought, it could be conjectured that the very way life operates on a given Earth-like planet may prevent life from existing on another Earth-like planet.

However, here comes the pseudo-retrocausality analysis past its usefulness, as it is stated that life exists on both planets, yet it is concluded that life could not arise on one planet. Furthermore, each planet is encompassed by an unconditional probability, i.e. a probability unaffected by the preceding of other events. Thus, what happens to an Earth-like planet at one end of the galaxy does not affect an Earth-like planet at another end.

So if one takes causality in the right order, this however still means that two Earth-like planets, even if they share the same initial conditions, subsequently entail different situations. So if life arises on two Earth-like planets, then the segments of these planets and their biodiversity will become more different as time passes, meaning that two present-day Earth-like planets entail an extreme improbability.

However, it is the case in probability theory that even the most improbable events, having a non-zero probability of happening, can occur. So while the fact that two present-day Earth-like planets, with regard to the order of the segments and biodiversity, are completely the same represents an extreme improbability, they can in principle indeed be completely the same. Thus, the following example can be forwarded.

There are 12 segments. It is assumed that the first segment S1 must exist on every Earth-like planet, and it is assumed for the sake of simplicity that chemical evolution will occur in the next segment S2 on each planet, so that there are subsequently 10 segments. A world's geology is assumed to proceed more predictably than life, so that each segment follows the same way on each planet, for the sake of the example, it can be set to $P(x) = ½$. Thus, we have the following:

$$P(x) = \left(\frac{1}{2}\right)^{10} \tag{1}$$

$$= 9.8 \times 10^{-4},$$

or less than one in one thousand, yet not zero. Thus, if 1000 other Earth-like planets are located, then one of them should be like the present-day Earth in terms of the order of the segments. Considering that the number of inhabited worlds in a galaxy such as the Milky Way has been estimated to be at least 300 million potentially habitable worlds [Bryson et al., 2021], this is a promising number in the search for other Earths.

However, if the calculation is made for biodiversity, the picture changes drastically. That a world's biology proceeds more unpredictably than its geology is the case, so the following assumptions are made here:

A bacterium survived being engulfed by an archaea with which the eukaryotic organism arose by symbiogenesis [Roger et al., 2017]. $P(x) = 1/1000$.

Some descendants of this eukaryotic organism joined together, leading to multicellular life [Grosberg and Strathmann, 2007]. $P(x) = 1/1000$.

Some of this multicellular life eventually acquired the ability to make rudimentary decisions. $P(x) = 1/1000$. Thus, we have the following:

$$P(x) = \left(\frac{1}{1000}\right)^3 \tag{2}$$

$$= 1 \times 10^{-9},$$

or less than one in one billion, yet still not zero. Thus, here is one past among the 300 million potentially habitable worlds. Furthermore, the assumptions and calculations are very simplified. Apart from the fact that the segments and, certainly, life follow dependent events, many other major and less major events in the history of terrestrial life are not included. Furthermore, the numbers are relatively arbitrary and very generous.

Yet, on this basis, it can already be seen that the probability of another present-day Earth-like planet in terms of the same segments and biodiversity becomes extremely improbable. With more complex calculations, the probability of finding another Earth-like planet entirely similar with regard to biodiversity in even the entire galaxy cluster is virtually non-existent.

Note that this does not refer to the emergence and continued presence of life on another present-day Earth, but only that it will reflect biodiversity as on present-day Earth.

It does apply, however, that the probability that two Earth-like planets in the past are the same with regard to life is much higher than that of two Earth-like planets being the same as each other in the present with regard to life. This result is due to life starting as the simplest functional organism imaginable. Thus, life can only arise as the most rudimentary bacteria-like organism possible [von Hegner, 2019; 2021], which means that unicellular life will have much in common on such initial Earth-like planets.

Naturally, many factors apply on an Earth-like planet. That life can arise and life can be maintained are two of the important ones. Thus, that life in the present-day segment originates from life in the previous segment differs from that life in the previous segment will necessarily lead to life in a later segment, meaning that the fact that life occurs on a planet does not imply that life will continue existing on that planet.

However, brushing that aside, unicellular life, if it arises, will have much in common on such initial Earth-like planets.

This is important in terms of the search for life on an Earth-like planet, because as soon as the simplest functional bacteria-like life form crossed the threshold of chemical evolution, life has persisted on this planet.

There is no direction or drive for the complexification of life in a strict Darwinian sense. The later segments may thus well solely have microbial life.

This is the basis for the Disparatis conjecture, stating that unicellular life is common in the galaxy, while multicellular life is rare in comparison [von Hegner, 2022].

So while multicellular life has arisen on Earth, this does not necessarily always do so on other planets. Indeed, the unicellular organism has remained the dominant life on the Earth. It has always been there and will likely remain there until the phases of the sun towards the red-giant stage end life on the Earth.

So here the Earth-like planet used as a search criterion for life elsewhere seems to be more justifiable, as there is a connection between an Earth-like planet and unicellular life.

However, while unicellular life is similar in terms of being single-celled, there is still a vast biodiversity between them on present-day Earth. Thus, that life exists on a planet such as present-day Earth, and that life has arisen on a planet such as the Earth of the past still entail two very different situations. Another Earth-like planet where life starts at the same time and in the same way will still be able to become more and more different as time goes on. Additionally, the segments, as discussed, may not geologically proceed in the same order on another world either.

However, this is life as we know it, but the picture may be more complicated than that. As mentioned, life arose relatively early in the Earth's history. Several lines of evidence point to traces of life as early as 4.1 billion years ago [Bell et al., 2015], in the segment that was at the end of the Hadean eon. However, the conditions in these segments differed from those in the later segments.

So for life on a world, one can fundamentally ask, which came first, the conditions appropriate for life or the life appropriate for the conditions? This is a chicken-or-egg kind of paradox.

By this is meant, did this particular type of life arise because it fit these conditions, and that another type of conditions could produce another type of life? That is, does life go beyond the *modus operandi* seen in this particular life?

So instead of asking what circumstances lead to the emergence of life, it may be more correct to ask what circumstances prevent life's *opus operatum* from arising.

If Earth-like life arose this way because the conditions suited it, then another Earth-like planet in other segments would be able to produce a different type of life.

That this life is not seen can be either because it cannot compete with the first life that destroys its possibility to arise, or because it exists but is so different that it has not yet been discovered.

Life on another Earth-like planet may not have to arise at the same time as on Earth, because although their first segments may be relatively similar, different factors, e.g. the early violent period in a solar system may eliminate that life so that it will only be able to arise in the later segments, and thus another type of life.

If this is the case, then this will make life on two Earth-like planets even more different from each other, and thus make a search criterion between life and Earth-like planets even harder to justify.

**6. Conclusion**

The analyses in this work have shown that the term Earth-like planet is more limited than assumed when first posited, as it does not represent a clear-cut entity. A general Earth-likeness cannot be abstracted and it is more accurately treated as a fuzzy world. This is partly due to the dynamic nature of the world itself, partly due to the variance of life itself, and finally due to the intermingled dynamic interaction between the environment and life, which means that there is a certain fuzziness with Earth-like planets. So even if an Earth-like planet exists in the habitable zone, which is one of the criteria for being Earth-like, this planet can appear as different worlds over time, and even Earth-like planets of the same age as present-day Earth can be quite different from this.

These analyses do not represent a limitation in the search for Earth-like planets. On the contrary, the plethora of an Earth-like planet shows that the possibility of such worlds can be greater than assumed, as there is not just one type of planet possible with life, but in fact, many. Thus, this is not a limitation of the search for life, but an enrichment of it.

However, although there is a plethora of types an Earth-like planet goes through, this is not the same as the principle of plenitude, as such a planet will not necessarily achieve all attainable types in its lifetime.

Yet still the possibility of inhabited worlds will be greater than assumed. This includes the fact that we have limited ourselves here to talking about Earth-like planets, but worlds other than the Earth may also be capable of harbouring life. Thus, some of the segments the Earth was divided into can likely also be applied to different worlds.

However, while these analyses do not represent a limitation in the search for Earth-like planets, this is a limitation in the search for life on such worlds. Thus, that life makes the surrounding environment less favourable for itself, which is ultimately the reason why there is a biosignature, entails different situations in that the varying amounts of life can entail different situations in the present-day and the past planet. Thus, one challenge of such a fuzzified world is that an atmosphere far from thermodynamic equilibrium with regard to life can be deceptive.

A fuzzy world is an Earth-like planet whose segments with regard to life change over time in such a way in relation to each other that it shows the effect of life depending on a certain time and certain factors that interact in a certain way, so that the effect of that life is not necessarily there at another time. The particular single entity that is abstracted as the present-day Earth-like planet emerges as a result of things standing in a certain way at a certain moment, which means that the Earth-like presentation is limited by its presenting life in only some of its segments. Thus, the presence of life can be indeterminate.

This work is not a discussion of the probability of the occurrence of chemical evolution on each planet, but a discussion of the probability of how biological evolution proceeds differently on each planet, even with the same initial conditions, although it can be discussed whether Earth-like planets will necessarily lead to life.

Thus, that two present-day Earth-like planets with regard to life are alike entails an extreme improbability. This should not be surprising, as evolution has a component of local necessity and a component of chance, which means there is a fuzziness about it, in that it is not possible to present a complete predictive framework for it.

One may question whether biological evolution works in the same way on other solar system bodies, even other Earth-like planets. However, what is evolution by natural selection? Given a population of individuals that has (a) reproduction; (b) variation; and (c) differential success linked to variation, then evolution will inevitably be in operation. The individuals most successful in the environment will become more prevalent in the population over time, as a result of being naturally selected [Levin et al., 2019].

Thus, even if life consists of different materials than what applies to terrestrial life, then life must still follow these three mechanisms, as it must necessarily pass on information to the next generation, and when it passes on information there is also variation in it, ultimately due to chemistry and when there is variation, some traits will also statistically become more prevalent over time.

The analyses show that a general Earth-likeness cannot be abstracted, which in many ways represents a novel work. Thus, there is much room for further research.

One is the formulation of a more mathematically oriented framework that can deal more comprehensively than the brief calculations provided with a fuzzy world's mix of causality, reciprocity, correlation and probability. An issue complicating all this is the dynamic interplay between teleomatic processes and teleonomic activities.

Another thing that needs to be addressed is that one has to specify which of the segments are at play when searching for life elsewhere, but how is that possible when the age of the segments and the presence of life on them are not necessarily connected.

Another thing that needs to be addressed is how long a ghost biosignature can be generated on a world.

The analyses show that the segment with the present-day Earth-like planet is no more true than the other segments, which has consequences for astrobiology and the search for life elsewhere in the galaxy. They thus provide an impetus for further work.